\documentclass[a4paper]{article}

\author{G. Gubbiotti\footnote{e-mail:
gubbiotti@mat.uniroma3.it}, $\quad$  D. Levi\footnote{e-mail:
decio.levi@roma3.infn.it}, $\qquad $
C. Scimiterna\footnote{e-mail:
scimiterna@fis.uniroma3.it} \\
 Dipartimento di Matematica e Fisica, Universit\`a degli Studi Roma Tre,\\ e Sezione INFN di Roma Tre,\\ Via della Vasca Navale 84, 00146 Roma (Italy)
 }\title{On Partial Differential and Difference Equations with  Symmetries Depending on Arbitrary Functions.}

\usepackage{cite, amssymb}
\usepackage{braket}
\usepackage{amsmath}
\usepackage{booktabs}
\usepackage{amscd}
\usepackage[Q=yes]{examplep}
\usepackage{multirow}
\usepackage{rotating}
\usepackage{array}
\usepackage{longtable}
\usepackage{pdflscape}
\def\lf{\left(}
\def\rg{\right)}
\def\lq{\left[}
\def\rq{\right]}

\renewcommand{\epsilon}{\varepsilon}

\numberwithin{equation}{section}
\def\bea{\begin{eqnarray}}
\def\eea{\end{eqnarray}}

\newcommand{\Fp}[1]{F^{(+)}_{#1}}
\newcommand{\Fm}[1]{F^{(-)}_{#1}}

\newtheorem{theorem}{Theorem}

\allowdisplaybreaks

\begin{document}
\maketitle
\begin{abstract}
In this note we present some ideas on when Lie symmetries, both point and generalized, can depend on arbitrary functions. We show on a few examples, both in partial differential and partial difference equations when this happens. Moreover we show that the infinitesimal generators of generalized symmetries depending on arbitrary functions, both for continuous and discrete equations, effectively play the role of  master symmetries. 
\end{abstract}
\begin{center}
{\it Work dedicated to  Jiri Patera and Pavel Winternitz 
on the occasion of their 80th birthday.}
\end{center}

%%%%%%%%%%%%%%%%%%%%%%
\section{Introduction}
%%%%%%%%%%%%%%%%%%%%%%
In a seminal work of two century ago Medolaghi \cite{m98}, following Sophus Lie results on ordinary differential equations \cite{Lie74,Lie}, proposed the following work program for Partial Differential Equations (PDE's):
\begin{enumerate}
\item Determine all different kinds of infinite groups of point transformations in 3 variables.
\item For each of the obtained groups determine the invariant second order equations.
\end{enumerate}
In the framework of this research Medolaghi got, among other equations,  the Liouville equation
\bea \label{l1}
u_{xt}(x,t)=e^{u(x,t)}.
\eea 
The symmetry algebra of the  Liouville equation~(\ref{l1}) is given by the vector fields
 \begin{gather} \label{l2}
 X(f(x)) = f(x) \partial_x - f_x(x)  \partial_u, \qquad Y(g(y))=g(y) \partial_y - g_y(y)  \partial_u,
 \end{gather}
 where $f(x)$ and $g(y)$ are arbitrary smooth functions of their argument and by $f_x(x)$ and $g_y(y)$ their first derivative.
The commutation relations of the vector fields (\ref{l2}) are 
\bea \lq  X\lf f \rg , X ( \tilde f ) \rq   &=& X\lf f \tilde{f}_x- \tilde{f} f_x \rg , \quad   \lq  Y\lf g \rg , Y\lf \tilde{g}\rg \rq   = Y\lf g\, \tilde{g}_y - \tilde{g} \, g_y \rg , \nonumber \\ && \lq  X\lf f\rg , Y\lf {g}\rg \rq   = 0. \label{l3}\eea
The algebra (\ref{l2}, \ref{l3}) is isomorphic to the direct sum of two Virasoro algebras. %We denote it $L = vir_x \oplus vir_y$.

We can find also $S$--integrable equations \cite{Calogero}, i.e. equations integrable by the {\it Spectral transform}, which have an infinite dimensional group of point symmetries. Classical examples are the 2+1 dimensional Kadomtsev-Petviashvili equation\cite{dklw1,dklw2,s82}, the 2+1 dimensional Toda \cite{lw91}, the three wave interaction in three dimensions \cite{mw89}.

In the Example 5.7 of \cite{olver} one can find the construction of the point and generalized symmetries for the nonlinear first order wave equation 
\bea \label{l4}
u_t = u \,u_x, \quad u=u(x,t).
\eea
In evolutionary form they  are given by the characteristic
\bea \label{l5}
Q=u_xF(x+tu,u,t+\frac{1}{u_x}, \frac{u_{xx}}{u_x^3}),
\eea
where the function $F$ is an arbitrary function of its arguments. In particular the last term corresponds to generalized symmetries as 
it depends of the second derivative of the field. Results by Shabat and Zhiber \cite{zs79} show that also the Liouville equation, being Darboux integrable,  has such kind of symmetries:
\bea \label{l6}
Q=(D_x+u_x) \,F(w,w_x, \cdots, w_{k_1x})+( D_y+u_y) \,G(\bar w, \bar w_y, \cdots, \bar w_{k_2 y}),
\eea
where the operators $D_x+u_x$ and $D_y+u_y$ are the Laplace operators, $k_1$ and $k_2$ are two positive integer numbers, $F$ and $G$ are two arbitrary functions of their arguments, $D_x$ and $D_y$ stands for the total derivative with respect to the $x$ and $y$ variables and $w=u_{xx}-\frac 1 2 u_x^2$ and $\bar w = u_{yy}-\frac 1 2 u_y^2$ are the lowest order integrals of the Liouville equation in the $x$ and $y$ direction.

So  the nonlinear wave equation (\ref{l4}) as well as the Liouville equation (\ref{l1}) admit generalized symmetries depending on arbitrary functions. Then Medolaghi program could be extended to the case of generalized symmetries. We here present some preliminary ideas on a possible solution of this program.

In the following  Section we will present some results on the construction of PDE's presenting generalized symmetries depending on arbitrary functions and relate them to Darboux integrable equations. Then in Section 3 we present the counterpart of the previous results in the discrete case.  At the end  we present few conclusive remarks and conjectures.

%%%%%%%%%%%%%%%%%%%%%%
\section{Factorizable differential operators, Darboux integrable equations and symmetries depending on arbitrary functions.}
%%%%%%%%%%%%%%%%%%%%%%

The result (\ref{l5}) can be in principle generalized to any  differential equation of the first order as the equation for the symmetries, be them point or generalized ones,  can be solved on the characteristics as was the case of the Hopf equation (\ref{l4}). 
Let consider, with no loss of generality,  the example of a general first order autonomous equation in two independent variables, %as was the case of (\ref{l4}), 
\bea \label{m1}
u_t=f(u,u_x),\quad u=u(x,t),
\eea
where $f$ is an arbitrary function of its arguments.
The symmetries are given by their characteristics $Q(x,t,u(x,t),u_x(x,t), \cdots, u_{kx}(x,t))$ and its determining equation is given by 
\bea \label{m2}
D_t Q -\left[\frac{\partial f}{\partial_u} Q + \frac{\partial f}{\partial_{u_x}}
D_x Q \right]\Bigg |_{u_t=f}=0, 
\eea
i.e. a PDE of the first order for the various components from which the function  $Q$ depends. Eq. (\ref{m2}) is a first order differential equation for $Q$ which can be solved on the characteristics and whose solutions provide $k+2$ symmetry variables. Than, as is in the case of (\ref{l5}), the symmetries of (\ref{m1}) are given by arbitrary functions of the $k+2$ symmetry variables. So any first order PDE, linear or nonlinear, will have generically point and generalized symmetries depending on arbitrary functions of the symmetry variables. 

We can easily show an interesting consequence of the existence of generalized symmetries depending on arbitrary functions. For the sake of concreteness we consider the symmetries of (\ref{l4}).  Let us consider the subcase of (\ref{l5}) when 
\bea \label{m3}
Q_f=u_x \left [f_1(x+tu)+f_2\left(t+\frac{1}{u_x}\right)+f_3 \left(\frac{u_{xx}}{u_x^3}\right) \right],
\eea
i.e. the Hopf equation (\ref{l4}) has a symmetry generator of the form:
\bea \label{m4}
\hat X_f = Q_f \partial_u
\eea 
with the functions $f_i, \, i=1,2,3$  analytic in their argument. When $f_2$ and $f_3$ are zero than (\ref{m4}) is the infinitesimal generator of point symmetries depending on an arbitrary function $f_1$; when $f_1$ and $f_3$ are zero than (\ref{m4}) is the infinitesimal generator of contact symmetries depending on an arbitrary function $f_2$; when $f_1$ and $f_2$ are zero than (\ref{m4}) is the infinitesimal generator of generalized symmetries depending on an arbitrary function $f_3$. If we take two of such generators, $\hat X_f$ and $\hat X_g$, where $f$ and $g$ are two different functions of the same argument, what can we say of their commutator? It has been proved by B\"acklund \cite{i} that as soon as the characteristic $Q$ depends on derivatives of order higher than the first one, the symmetry group is infinite. The last symmetry group which can be finite is the one of the contact symmetries, if no arbitrary function is present. As was shown in the case of the Liouville equation the presence of a symmetry generator of point symmetries depending on an arbitrary function provide an infinite dimensional Lie algebra of point symmetries. 
%This is also the case for the Hopf equation by choosing $f_2=f_3=0$. 

Let us  carry out the the commutation of $\hat X_f$ and $\hat X_g$, for $f$ and $g$ equal to $f_1$ we get:
\bea \label{m5}
\left [ \hat X_{f_1}, \hat X_{g_1} \right]=\hat X_{f_1 g_1^{'}-f_1^{'} g_1},
\eea 
equal to  $f_2$ we get:
\bea \label{m6}
\left [ \hat X_{f_2}, \hat X_{g_2} \right]=0,
\eea 
and equal to $f_3$ we get:
\bea \label{m7}
\left [ \hat X_{f_3}, \hat X_{g_3} \right]=\left[f_3^{''}g_3^{'}-f_3^{'} g_3^{''}\right] \frac{(u_{xxx}u_x - 3 u_{xx}^2)^2}{u_x^9}\partial_u.
\eea 
The algebra (\ref{m5}) turn out to be  a Virasoro algebra of Lie point symmetries. The infinitesimal generators of the contact symmetries (\ref{m6}) commute but the infinitesimal generator depending on an arbitrary function of the generalized symmetry take the form of a {\it master symmetry} as the commutator of two such symmetries provide a symmetry of higher order (\ref{m7}).

The existence of arbitrary functions of generalized symmetries is not limited to the case of first order differential equations. It can easily extended to higher order PDE's when the differential operator which define the equation is {\it factorizable}. Few authors  \cite{t00,ja} showed through the Laplace cascade method that factorizable linear PDE's are Darboux integrable if the cascade terminates.  We can show that a factorizable PDE admits as a subclass of symmetries the symmetries of its first order PDE. As an example let us consider  the case  of second order factorizable PDE's.
Let us consider the second order autonomous partial differential equation for one dependent variable $u$ in the two independent variables $x$ and $t$
\bea \nonumber 
&&E_1=u_{tt} + \left[g(u, u_x)-f(u,u_x)_{u_x}\right]u_{xt}-f(u,u_x)_{u_x} g(u,u_x) u_{xx} - \\ && \qquad -f(u,u_x)_u \left[ u_t - g(u,u_x) u_x \right]= 0, \label{m8}
\eea
where $f$ and $g$ are two arbitrary functions of their arguments.
Defining the  first order autonomous PDE for $u=u(x,t)$
\bea \label{m8a}
E_0=u_t-f(u,u_x)=0
\eea
 (\ref{m8}) is factorizable as:
\bea \label{m9}
E_1=\left[ \partial_t + g(u,u_x) \partial_x \right] E_0=0.
\eea
The symmetries of $E_0$ are given by the infinitesimal generator
\bea \label{m10}
\hat X= Q_0(x,t,u,u_x,u_t, u_{xx}, \cdots) \partial_u
\eea
and the determining equation is written as
\bea \label{m11}
D_t Q_0 -f(u,u_x)_u Q_0 - f(u,u_x)_{u_x} D_x Q_0\bigg |_{E_0=0}=0 
\eea
where $D_t Q_0$ and $D_x Q_0$ are the coefficient of the prolongation of $\hat X$ with respect to $u_t$ and $u_x$, $D_t$ and $D_x$ being the total derivatives with respect to its index. The determining equation for the symmetries of $E_1=0$  of infinitesimal generator 
\bea \label{m10a}
\hat Y= Q_1(x,t,u,u_x,u_t, u_{xx}, \cdots) \partial_u
\eea
are given by
\bea \label{m12}
\mbox{pr}\,\hat Y E_1\bigg |_{E_1=0}=0. 
\eea
Eq. (\ref{m12}) is complicate as it requires the application of the second prolongation  of $\hat Y$, but it is a straight  forward calculation. So we do not write it down here. By a direct calculation it is easy to show that (\ref{m12}) is satisfied by the solution of (\ref{m11}) when $E_0=0$ is satisfied. This shows that the second order autonomous factorizable PDE $E_1=0$ admits the same symmetries as $E_0=0$ and thus it can have arbitrary function dependent generalized symmetries as $E_0=0$ does. This proof can be extended to PDE's of any order and of any number of variables. The relation of this factorization to Laplace cascade method and Darboux integrability is still to be understood.

%%%%%%%%%%%%%%%%%%%%%%
\section{Discrete equations with generalized symmetries depending on an arbitrary function.} 
%%%%%%%%%%%%%%%%%%%%%%
Partial Difference Equations (P$\Delta$E's) can have have symmetries depending on arbitrary functions. The presence of arbitrary functions of point symmetries has been shown to appear, as in the continuous case \cite{ba,i,gsw,s04}, when the P$\Delta$E is linearizable \cite{ls}. Here we show on some examples that one can find  P$\Delta$E's which have generalized symmetries depending on arbitrary functions. When they appear the system is usually  linearizable  but a complete theory in this case is absent. We have some results on Darboux integrable discrete equations \cite{st,s14,gy,vere,as, gy12} which correspond to  P$\Delta$E's which have two distinct conserved quantities in the two different directions of the discrete plane. The complete classification of Darboux integrable equations on the lattice is absent and only some simple classes are worked out in the references mentioned above. Darboux integrable equations turn out to be  linearizable but the other way around is not true.

Here in the following we presents  three linearizable P$\Delta$E's \cite{h,gsl} which have generalized symmetries depending on an arbitrary function. Then, in a subsection, we present the case of the completely discrete Liouville equations. The first two equations are quad graph equations   which do not possess the tetrahedron property. They turn out to be  linearazable and have  generalized symmetries  of any order \cite{h}. Then we consider an equation of the Boll classification, $_t H_1^{(\epsilon)}$, which is non autonomous, linearizable  and has three point generalized symmetries \cite{gsl}. 

The three equations  are:
\begin{enumerate}
\item The first equation  belongs to  the classification of compatible equations around the cube with no tetrahedron property presented by Hietarinta in \cite{h}:
\bea \label{n1}
v_{m+1,n} v_{m,n+1}+v_{m,n}v_{m+1,n+1}=0.
\eea
The three point symmetries in the $m$--direction of (\ref{n1}) are given by the characteristic
\bea \label{n2}
Q_{m,n}=v_{m,n} F\left((-1)^n \frac{v_{m+1,n}}{v_{m,n}}, (-1)^n\frac{v_{m,n}}{v_{m-1,n}}\right).
\eea
%with the condition $F_m(x,y)=F_m(-x,-y)$.
As (\ref{n1}) is symmetric in the exchange of $n$ and $m$ the generalized symmetry (\ref{n2}) is valid also the direction $n$ with the role of $n$ and $m$ interchanged. 

We can easily find two conserved quantity 
\bea \label{n2aa}
W_1=(-1)^n \frac{v_{m+1,n}}{v_{m,n}}, \qquad W_2=(-1)^m \frac{v_{m,n+1}}{v_{m,n}}
\eea
 such that 
\bea \label{nb2}
&(T_2 - 1)W_1 = 0, \quad &(T_1 - 1)W_2 = 0, \\ \nonumber  &T_2h_{m,n} =h_{m,n+1}, \quad &T_1h_{m,n} = h_{m+1,n}.
\eea Then (\ref{n1}) is Darboux integrable equation and (\ref{n2}) is just the three point subcase of a general characteristic in the $m$ direction
\bea \label{na2}
Q_{m,n}=v_{m,n} F\left((-1)^nT_{1}^{-N}\frac{v_{m+1,n}}{v_{m,n}},\ldots,(-1)^nT_{1}^{M}\frac{v_{m+1,n}}{v_{m,n}}\right),
\eea
with $N$ and $M$  two positive arbitrary integers such that  the number of points involved in the symmetry is given by $N+M+2$. A similar characteristic exists also in the $n$ direction with the role of $n$ and $m$ interchanged.
%Eq. (\ref{n1}) is a subcase of the first equation in the List 4 of Darboux integrable equations presented in \cite{gy}. Its integrals are:\bea \label{n2a}&&(T_1 - 1)W_2 = 0, \quad (T_2 - 1)W_1 = 0,\\  \nonumber&&W_1=\frac{(u_{m+2,n}-u_{m,n})(u_{m+1,n}-u_{m-1,n})}{(-1)^{n}}, \\ \nonumber&&W_2=\left( \frac{i u_{m,n}}{u_{m,n+1}}\right )^{(-1)^m},\eeawhere $T_1h_{m,n} = h_{m+1,n}$ and  $T_2h_{m,n} =h_{m,n+1}$. 

Eq. (\ref{n1}) is linearizable in three ways:
\begin{itemize}
\item By the point transformation $v_{0,0}\doteq e^{x_{0,0}}$, so that $x_{0,0}=\log v_{0,0}$ (without loss of generality, $\log$ can be always taken to stand for the principal value of the complex logarithm), is transformed into the linear equation $x_{0,0}-x_{1,0}-x_{0,1}+x_{1,1}=i\pi$.
\item By the Hopf--Cole transformation $v_{1,0}/v_{0,0}\doteq w_{0,0} \; (v_{0,0}\not=0)$ into the ordinary difference equation $w_{0,1}+w_{0,0}=0$.
\item By the Hopf--Cole transformation $v_{0,1}/v_{0,0}\doteq t_{0,0} \; (v_{0,0}\not=0)$ into the ordinary difference equation $t_{1,0}+t_{0,0}=0$.
\end{itemize}

\item A second equation which  belongs to  the classification of compatible around the cube with no tetrahedron property presented by Hietarinta in  \cite{h} is:
\bea\label{n3}
&&v_{m,n}+v_{m+1,n}+v_{m,n+1}+v_{m+1,n+1} + v_{m+1,n}v_{m,n+1}v_{m+1,n+1}+\\ \nonumber &&\qquad +v_{m,n}\left[ v_{m+1,n} v_{m,n+1}+v_{m+1,n}v_{m+1,n+1}+v_{m,n+1}v_{m+1,n+1}\right]=0
\eea
whose three point symmetries in the $m$--direction are given by
\bea \label{n4}
&&Q_{m,n}=(-1)^n(1-v_{m,n}^2) K\left[\left(\frac{(1-v_{m,n})(1-v_{m-1,n})}{(1+v_{m,n})(1+v_{m-1,n})}\right)^{\left(-1\right)^n},\right. \\ \nonumber &&\left. \qquad \qquad \qquad \qquad \qquad \qquad ,\left(\frac{(1-v_{m,n})(1-v_{m+1,n})}{(1+v_{m,n})(1+v_{m+1,n})}\right)^{\left(-1\right)^n}\right].
\eea
Eq. (\ref{n3}) admits a  fourfold discrete symmetry  given by
\bea
\label{f1} v_{0,0}\rightarrow v_{0,0}^{\left(\epsilon\right)}\doteq\frac{\left(1+\epsilon\right)v_{0,0}+1-\epsilon}{\left(1-\epsilon\right)v_{0,0}+1+\epsilon},
\eea
where $\epsilon$ is one of the four quartic roots of unity, $\epsilon=(\pm i, \,\pm 1)$.
Eq. (\ref{n3}) is symmetric in the exchange of $n$ and $m$ and consequently the generalized symmetry (\ref{n4}) is valid also in the direction $n$ interchanging the role of $n$ and $m$. It  is not contained in the lists of Darboux integrable discrete equations  \cite{gy} and  \cite{st},  however we can prove that it is Darboux integrable as it admits the following integrals which satisfy (\ref{nb2}):
\bea \label{n3a}
&&W_1=\left(\frac{\left(1-v_{m,n}\right)\left(1-v_{m+1,n}\right)}{\left(1+v_{m,n}\right)\left(1+v_{m+1,n}\right)}\right)^{\left(-1\right)^n},\\
\nonumber &&W_2=\left(\frac{\left(1-v_{m,n}\right)\left(1-v_{m,n+1}\right)}{\left(1+v_{m,n}\right)\left(1+v_{m,n+1}\right)}\right)^{\left(-1\right)^m}.
\eea
The symmetry characteristic (\ref{n4}) is nothing but the reduction to three points of a general case depending on $N+M+2$ points with $N$ and $M$ arbitrary positive integers,
\bea \label{nn4}
&&Q_{m,n}=(-1)^n \,(1-v_{m,n}^2) K\left[T_1^{-N}\left(\frac{(1-v_{m,n})(1-v_{m-1,n})}{(1+v_{m,n})(1+v_{m-1,n})}\right)^{\left(-1\right)^n},\cdots, \right. \\ \nonumber &&\left . \qquad \qquad \qquad \qquad \cdots, T_1^{M}\left(\frac{(1-v_{m,n})(1-v_{m-1,n})}{(1+v_{m,n})(1+v_{m-1,n})}\right)^{\left(-1\right)^n} \right].
\eea
 A similar characteristic exists also in the $n$ direction with the role of $n$ and $m$ interchanged.

Eq. (\ref{n3}) is linearizable by the  point transformation:
\bea \label{n4a}
v_{m,n}\doteq\frac{1+e^{x_{m,n}}}{1-e^{x_{m,n}}}
\eea
into the linear equation 
\bea \label{f2}
x_{m,n}+x_{m+1,n}+x_{m,n+1}+x_{m+1,n+1}=2iz\pi,
\eea
 where $z=-1$, $0$, $1$, $2$.  The indeterminacy of the inhomogeneous term of (\ref{f2}) takes into account the fourfold discrete symmetry (\ref{f1}) of (\ref{n3}). It is worth while to notice that  inverting (\ref{n4a}) we get $x_{0,0}=\log\frac{v_{0,0}-1}{v_{0,0}+1}$, where
without loss of generality, the function $\log$ can always be taken as the
principal value of the complex logarithm \cite{sl13}.
\item $_t H_1^{(\epsilon)}$ is:
\bea \label{eql}
&&\left(x_{m,n}-x_{m+1,n}\right)\left(x_{m,n+1}-x_{m+1,n+1}\right)-\\ \nonumber  && \qquad \epsilon^2\alpha_{2}\left({ F}_{n}^{\left(+\right)}x_{m,n+1}x_{m+1,n+1}+{ F}_{n}^{\left(-\right)}x_{m,n}x_{m+1,n}\right)-\alpha_{2}=0.
\eea
where ${ F}_{n}^{\left(\pm \right)}=\frac{1 \pm (-1)^n}{2}$ and $\epsilon$ is an arbitrary constant as well as $\alpha_2$.
Eq. (\ref{eql}) is Darboux integrable, as we can find two integrals in the $m$ and $n$ directions which satisfy (\ref{nb2}). The $m$-integral being given by
\bea
\label{f3} &&W_{2}\doteq\Fp{n} s+\Fm{n}(T_2+1)u,\\
\nonumber &&s\doteq\frac{x_{m,n+1}-x_{m,n-1}}{1+\epsilon^2x_{m,n+1}x_{m,n-1}},\ \ \  u\doteq x_{m,n}-x_{m,n-1}.
\eea 
  The $n$ direction integral is different, as $_t H_1^{(\epsilon)}$ is not symmetric in the exchange of $n$ and $m$, and is given by
\bea
\label{f4} &&W_{1}\doteq\alpha \left(T_{2}+I\right)\Fp{n} v +\beta \left(T_{2}+I\right)\Fm{n} t,\\ \nonumber && t\doteq\frac{x_{m,n}-x_{m-1,n}}{1+\epsilon^2x_{m-1,n}x_{m,n}}, \ \ \  v\doteq x_{m,n}-x_{m-1,n}
\eea
\noindent where $\alpha$  and $\beta$ are two arbitrary constants. Eq. (\ref{f4}) are effectively two  integrals  as $\alpha$  and $\beta$ are two independent constants.

In \cite{gls} we  constructed its three point generalized symmetries along the direction $m$:
{\scriptsize\bea
&&Q_{m,n}=\Fp{n}\left\{\frac{\alpha_{2}\left(v^2+\epsilon^2\alpha_{2}^2\right)}{\left(\bar v-v\right)\left(\bar v+v\right)}B_m\left(\frac{\alpha_{2}}{\bar v}\right)-\frac{\alpha_{2}\left(\bar v^2+\epsilon^2\alpha_{2}^2\right)}{\left(\bar v-v\right)\left(\bar v+v\right)}B_{m-1}\left(\frac{\alpha_{2}}{v}\right)+\right.\label{Sator1}\\
\nonumber && \left. +\left[x_{m,n}-\frac{\left(\bar v^2+\epsilon^2\alpha_{2}^2\right)v}{\left(\bar v-v\right)\left(\bar v+v\right)}\right]\omega+\gamma_{n}\right\}+\Fm{n}\left[\frac{\bar t^2t^2}{\left(\bar t-t\right)\left(\bar t+t\right)}\left(B_m\left(\bar t\right)-B_{m-1}\left(t\right)\right)-\right.\\
\nonumber && \left. \qquad -\frac{\bar t^2t}{\left(\bar t-t\right)\left(\bar t+t\right)}\omega+\delta_{n}\right]\left(1+\epsilon^2x_{m,n}^2\right), 
\eea}%
where $B_m\left(y\right)$, $\gamma_n$ and $\delta_n$ are generic functions of their arguments, $\omega$ is an arbitrary constant and  $\bar v=T_1 v$ and $\bar t=T_1 t$.
 Let us  note that any free function and free parameter may eventually depend on $\alpha_{2}$ and $\epsilon$. It is possible to demonstrate that, as long as $\epsilon\not=0$, no $n-$independent reduction of the above symmetry exists.

The three-points generalized symmetry along the direction $n$ is different as the equation is not symmetric. It is:
\begin{equation}
    \begin{aligned}
        Q_{m,n} &=  \Fp{n}\, \left( 
        {B}_{n} \left( s
     \right)+{  \kappa}_{{n}} \right) 
     \\
     &+{  \Fm{n}} \left( 1+
     {\epsilon}^{2}x_{{m,n}}^{2} \right) \left({C}_{n} \left( \bar u+u \right) +{  \lambda}_{{n}} \right),
    \end{aligned}
    \label{eq:symm_m}
\end{equation}
where $B_{n}(y)$ and $C_{n}(y)$ are arbitrary functions of their argument  and of the
lattice variable $n$, $\bar u = T_2 u$
  and $\kappa_n$ and $\lambda_n$ are arbitrary functions of the lattice variable $n$. Due to the complexity of the three point generalized symmetries we are not able, in this case,  to evince the Laplace operators and thus write the generalized symmetries depending on an arbitrary function for any number of points.\end{enumerate}
As in the case of PDE's presented in the previous Section also here the generalized symmetry depending on  an arbitrary function play the role of a master symmetry.  Let us consider , as an example, the commutation of two generalized symmetries of (\ref{n1}) of infinitesimal generators $\hat X_{F_m}=v_m F_m\partial_{v_m}$ and $\hat X_{G_m}=v_m G_m\partial_{v_m}$ characterized by the arbitrary functions $F_m=F\left( (-1)^n\frac{v_{m+1}}{v_m}\right)$ and $G_m=G\left((-1)^n \frac{v_{m+1}}{v_m}\right)$ corresponding to (\ref{na2}) with $N=M=0$. We have:
\bea \label{n5}
\left [ \hat X_{F_m}, \hat X_{G_m} \right]=(-1)^n v_{m+1} H_m\partial_{v_m}.
\eea 
where 
\bea \label{n6}
H_m= \left[ \Delta F_m G_{m}' - \Delta G_m F_{m}'\right] .
\eea
In (\ref{n6}) $F_{m}'$ means the derivative of $F_m$ with respect to its argument and the operator $\Delta$ is such that $\Delta F_m = F_{m+1}-F_m$. The function $H_m$ depends on more points than the functions $F_m$ and $G_m$ and so the generator $\hat X_{F_m}$ plays the role of a master symmetry.

%%%%%%%%%%%%%%%%%%%%%%%
\subsection{The discrete algebraic Liouville equations}

\noindent Among the many different completely discrete Liouville equations which go in the continuous limit in the algebraic Liouville equation
\bea \label{pippo}
v v_{xy} -v_x v_y = v^3,
\eea 
obtained from (\ref{l1}) by setting $v=e^u$, let's mention the following four:
\begin{itemize}
\item The Tzitzeica-Liouville equation, given in \cite{Ad}:
\bea \nonumber
h_{m,n}h_{m+1,n+1}\left(h_{m+1,n}-1\right)\left(h_{m,n+1}-1\right)-\left(h_{m,n}-1\right)\left(h_{m+1,n+1}-1\right)=0;
\eea
\item The potential Hirota-Liouville equation, given in \cite{s14, H1}:
\bea \nonumber
v_{m,n}v_{m+1,n+1}-v_{m+1,n}v_{m,n+1}+1=0;
\eea 
\item The Hirota-Liouville equation, given in \cite{s14, H2}:
\bea \nonumber
u_{m,n}u_{m+1,n+1}-\left(u_{m+1,n}-1\right)\left(u_{m,n+1}-1\right)=0;
\eea
\item The Adler-Startsev Liouville equation, given in \cite{as}:
\bea
t_{m,n}t_{m+1,n+1}\left(1+\frac{1}{t_{m+1,n}}\right)\left(1+\frac{1}{t_{m,n+1}}\right)-1=0.\label{Camese}
\eea
\end{itemize}
These P$\Delta$E's can be transformed one into the other by the following  transformations:
\bea
u_{m,n}&=&\frac{h_{m,n}}{h_{m,n}-1},\ \ \ h_{m,n}\not=1,\nonumber\\
u_{m,n}&=&v_{m+1,n}v_{m,n+1},\ \ \ \ \ \ \ \ \ \ \ \ \ \ \ \ \nonumber\\
u_{m,n}&=&-\frac{1}{t_{m,n}}.\ \ \ \ \ \ \ \ \ \ \ \ \ \ \ \ \ \ \,\nonumber
\eea
\noindent The generalized symmetries of (\ref{Camese}) along the direction $m$ are given $\forall p\in\mathcal{Z}$ and $\forall \, N\in\mathcal{N}_{0}$ by
\bea 
\frac{dt_{m,n}}{d\epsilon}=-\left(1+t_{m,n}\right)\left(T_{1}-1\right)\left[\left(\frac{t_{m,n}}{1+\frac{t_{m,n}\left(1+t_{m-1,n}\right)}{t_{m-1,n}}}\right) f\left(T_{1}^{p}W_1,\ldots,T_{1}^{p+N}W_1\right)\right],\label{s1}
\eea
\noindent where $\epsilon$ is the group parameter,  $f\left(x,y,\ldots,z\right)$ is an arbitrary function of its arguments and $W_1$ is the Darboux integral of the Liouville equation (\ref{Camese}) along the direction $n$, given in \cite{as} by
\bea
W_1=\left(1+\frac{t_{m,n}\left(1+t_{m-1,n}\right)}{t_{m-1,n}}\right)\left(1+\frac{t_{m,n}}{t_{m+1,n}\left(1+t_{m,n}\right)}\right),\nonumber
\eea 
\noindent satisfying (\ref{nb2}). Eq. (\ref{s1}) is the equivalent of formula (\ref{l6}) introduced by  Zhiber and Shabat for the continuous Liouville equation. 

Choosing $p=-1$ and $N=1$ in (\ref{Camese}) we obtain the most general generalized symmetry depending at most on the five points $t_{m-2,n}$, $t_{m-1,n}$, $t_{m,n}$, $t_{m+1,n}$ and $t_{m+2,n}$
\bea
\frac{dt_{m,n}}{d\epsilon}=-\left(1+t_{m,n}\right)\left(T_{1}-1\right)\left[\left(\frac{t_{m,n}}{1+\frac{t_{m,n}\left(1+t_{m-1,n}\right)}{t_{m-1,n}}}\right) g\left(T_{1}^{-1}W_1,W_1\right)\right].\label{s2}
\eea
 If $\partial_{x}g\left(x,y\right)\not=0$ and $\partial_{y}g\left(x,y\right)\not=0$, then we get a  five-points symmetry; if $\partial_{x}g\left(x,y\right)=0$ and $\partial_{y}g\left(x,y\right)\not=0$, then we get a  four-points symmetry depending on the asymmetric set $t_{m-1,n}$, $t_{m,n}$, $t_{m+1,n}$ and $t_{m+2,n}$; if $\partial_{x}g\left(x,y\right)\not=0$ and $\partial_{y}g\left(x,y\right)=0$, then we get a  four-points symmetry depending on the asymmetric set $t_{m-2,n}$, $t_{m-1,n}$, $t_{m,n}$ and $t_{m+1,n}$; finally if $g=1$ we get the three-points symmetry
\bea
\frac{dt_{m,n}}{d\epsilon}=-\left(1+t_{m,n}\right)\left(T_{1}-1\right)\left[\frac{t_{m,n}}{1+\frac{t_{m,n}\left(1+t_{m-1,n}\right)}{t_{m-1,n}}}\right],\label{s3}
\eea
\noindent which doesn't depend on any arbitrary function. As it is evident from (\ref{Camese}) there is no way that we can get a Lie point symmetry in agreement with the results presented in \cite{lmw1,lmw2}. The lowest possible symmetry is a generalized symmetry depending on three points (\ref{s3}). 
Eq. (\ref{Camese}) is symmetric in the exchange of the $m$ and $n$ indices. So the symmetries in the $n$ directions are trivially given by (\ref{s1}) with $T_1$ substituted by $T_2$ and the shifts in $m$ substituted by shifts in $n$.

%%%%%%%%%%%%%%%%%%%%%%
\section{Conclusive remarks.}
%%%%%%%%%%%%%%%%%%%%%%
In this note we presented a set of results partially distributed in a series of articles of different authors concerning the presence of symmetries depending on arbitrary functions both in the continuous and in the discrete setting. Few of the words associated to these systems are {\it Darboux integrable systems}, {\bf linearizable systems}, {\it factorizable differential operators} but not all necessarily proved to be  connected to the presence of symmetries depending on arbitrary functions. Darboux integrable systems in two independent variables are systems, studied primarily by G. Darboux, characterized by the presence of at least a couple of conserved quantities in the two independent variables. As far as it is known in the literature Darboux integrable systems are linearizable and the primary example is the Liouville equation (\ref{l1}) which has also arbitrary function dependent  symmetries, both point and generalized \cite{zs}. 

Here we showed that generically first order PDE's have symmetries depending on arbitrary functions as the symmetry determining equations are characterized by just first order PDE's which can be solved on the characteristics in term of arbitrary functions of the symmetry variables. These same symmetries can be found in PDE's of higher order when the differential operator is factorized in the product of lower order ones which, at the bottom, is just a first order one. Tsarev in \cite{t00} correlate Darboux integrable systems with factorizable differential operators. 

The situation is less clear in the discrete setting. The classification of Darboux integrable systems is not complete for P$\Delta$E's even on the square graph.  Few results \cite{vere,as} are known on factorizable difference operators in the framework of Darboux integrable systems and   symmetries.
 Nothing has been done up to now to characterize partial difference equations whose symmetries are written in term of arbitrary functions of symmetry variables. 
 %As a final observation we can notice that the generalized symmetries (\ref{s1}) and its reductions (\ref{s2}) and (\ref{s3}) may not be symmetric in the exchange of $m-k$ in $m+k$ for any $k$ involved in them. This is an indication that some of the symmetries are not $S$--integrable differential difference equations \cite{y,ly}.
Moreover, let us notice  that the generalized symmetries we obtain for Darboux integrable
equations as (\ref{na2}, \ref{nn4}, \ref{s1}) do not define, in general, $S$--integrable differential difference
equations. Indeed these symmetries  do not always satisfy the necessary condition for the $S$--integrability, that the highest order shift
in the lattice variable is the opposite of the lowest  one  \cite{y,ly}.

As a result of this note we can conjecture the following  theorem:
\begin{theorem} Necessary and sufficient conditions for a PDE and a P$\Delta$E to have symmetries, possibly point but surely generalized, depending on arbitrary functions is that the system be Darboux integrable.
\end{theorem}

Some of the results presented here are new and never presented anywhere. Among them let us mention the structure of the Lie algebra of the symmetries of the Hopf equation, the fact that both for PDE's and P$\Delta$E's generalized symmetries characterized by arbitrary functions provide master symmetries. Moreover the calculations of the generalized symmetries, the Darboux integrals and the linearizing transformation for the two nonlinear quad graph equations introduced by Hietarinta are new here as well as the generalized symmetries for the completely discrete Liouville equation and the Darboux integrals for $_t H_1^{(\epsilon)}$.

\section*{Acknowledgment}

\indent CS and DL  have been partly supported by the Italian Ministry of Education and Research, 2010 PRIN {\it Continuous and discrete nonlinear integrable evolutions: from water waves to symplectic maps}.

\noindent GG and DL are supported   by INFN   IS-CSN4 {\it Mathematical Methods of Nonlinear Physics}.

\end{document}